\documentclass[%
aps,pre,amsmath,amssymb,showpacs,preprint,longbibliography,
tightenlines
]{revtex4-2}

\usepackage{lineno}
\newcommand*\patchAmsMathEnvironmentForLineno[1]{
  \expandafter\let\csname old#1\expandafter\endcsname\csname #1\endcsname
  \expandafter\let\csname oldend#1\expandafter\endcsname\csname end#1\endcsname
  \renewenvironment{#1}
     {\linenomath\csname old#1\endcsname}
     {\csname oldend#1\endcsname\endlinenomath}}
\newcommand*\patchBothAmsMathEnvironmentsForLineno[1]{
  \patchAmsMathEnvironmentForLineno{#1}
  \patchAmsMathEnvironmentForLineno{#1*}}
\AtBeginDocument{
\patchBothAmsMathEnvironmentsForLineno{equation}
\patchBothAmsMathEnvironmentsForLineno{align}
\patchBothAmsMathEnvironmentsForLineno{flalign}
\patchBothAmsMathEnvironmentsForLineno{alignat}
\patchBothAmsMathEnvironmentsForLineno{gather}
\patchBothAmsMathEnvironmentsForLineno{multline}
}

\usepackage{mathtools}
\usepackage{color}
\usepackage{bigints}
\usepackage{multirow}
\usepackage{graphicx}
\usepackage{dcolumn}
\usepackage{bm}
\usepackage{subfigure}
\usepackage[dvipsnames]{xcolor}
\usepackage[colorlinks=true,linkcolor=purple,citecolor=Green,urlcolor=magenta]{hyperref}
\usepackage{multirow}
\usepackage{mathtools}

\newif\iffigure
\figurefalse
\figuretrue
\allowdisplaybreaks[3]
\usepackage{cprotect}

\def\SI{Supplemental Material}

\def\dd{\mathrm{d}}
\def\all{\mathrm{all}}

\def\um{\textmu m\relax}

\def\Ttrap{T_{\mathrm{trap}}}
\def\Trelease{T_{\mathrm{release}}}
\def\Tjitter{T_{\mathrm{jitter}}}
\def\mPIV{\textmu PIV}
\def\mPTV{\textmu PTV}
\def\TP{\mathrm{TP}}
\def\TO{\mathrm{TO}}

\def\bvx{\bar{v}_x}

\usepackage {cancel}

\usepackage{changes}

\def\figwidth{0.8}

\begin{document}

\preprint{\today, ver.~\number\time}

\newcommand{\titleA}{
}

\newcommand{\titleB}{
Optically-Trapped Particle Tracking Velocimetry
}

\title{
\titleB
}

\author{Tetsuro Tsuji}
\email{tsuji.tetsuro.7x@kyoto-u.ac.jp; corresponding author}
\author{Shoma Hashimoto}%
\author{Satoshi Taguchi}%
\affiliation{%
Graduate School of Informatics, Kyoto University, Kyoto 606-8501, Japan
}%

\date{\today\\\vspace{2em}}

\begin{abstract}
\noindent 
In this paper, we propose a microflow velocimetry based on particle tracking with the aid of optical trapping of tracers, namely, \textit{optically-trapped particle tracking velocimetry} (ot-PTV). The ot-PTV has two phases: a trap phase, in which individual tracers are trapped by an optical force and held at a measurement position; a release phase, in which the tracer is released and advected by the fluid flow, without interference from the optical force. The released tracer is subsequently trapped again by the optical force. By repeating the set of trap and release phases, we can accumulate the sequential images of the tracer that have the same initial position. Although the data acquisition rate of ot-PTV is lower than that of standard micro-resolution particle image velocimetry (\textmu PIV) due to the nature of pointwise measurement, an advantage of ot-PTV is that the measurement positions can be chosen by the experimenter. That is, even when tracers are scarce in a test section because of some external effects and/or the small size of the test section---ill-suited cases for \textmu PIV since the experimenter has to wait for tracers to diffuse into the test section (i.e., diffusion-limited situation)---ot-PTV remains efficient. The concept of ot-PTV is validated using a benchmark experiment, i.e., a pressure-driven flow in a straight microchannel with a square cross-section. An application to thermally-induced microflows is also demonstrated, where tracers can be scarce in a test section due to thermophoresis.
\end{abstract}

\maketitle
\titlepage
\thispagestyle{empty}


\section{\label{sec:intro}Introduction}
Flow velocimetry is a fundamental tool to investigate the characteristics of fluid flow fields. In microflows, where the spatial dimension is microscale or even down to nanoscale, conventional yet effective methods are micro-resolution particle image velocimetry (\mPIV) \cite{Santiago1998,Lindken2009} and micro-particle tracking velocimetry (\mPTV) \cite{Park2005}. These methods are successful in measuring microflows and have been further developed in various manners, e.g., by extending \mPTV~to a three-component measurement \cite{Lindken2006}. 

In particle tracking velocimetry (PTV), we acquire the sequential images of individual particles. Then, the displacements between the images, divided by a frame interval, are computed as the tracer velocities. 
Although the spatial resolution of PTV can be as fine as the tracer diameter, the noise inherent in the Brownian motion of tracers may be problematic, for instance, in the flow measurement of inhomogeneous creeping flows near boundaries \cite{Tsuji2023} or highly-localized flows around point-like heat sources \cite{Bregulla2016,Fraenzl2022}. 
To eliminate the effect of noise, ensemble averaging over many samples (i.e., the displacement data) is a standard approach. However, measurement positions in PTV are determined by tracer positions, which are not controllable. Therefore, to suppress the effect of noise sufficiently, the accumulation of the displacements at the same measurement position is necessary, which becomes increasingly difficult as the regions of interest become smaller and/or the flow becomes more localized.

A typical problematic case is found in a pioneering experimental study~\cite{Bregulla2019}, where the thermally-induced microflows around heated Janus microparticles (diameter 1~\um~or 8~\um) were investigated. Because of the small size of the test section, i.e., the region near the microparticle, the authors needed seven hours for a single trial, waiting for Brownian tracers to diffuse into the test section. Such a case can be classified as ``diffusion-limited", that is, experimenters have to wait for tracers until they diffuse into the region of interest. Another problematic case arises when external forces act on the tracers, pushing them out of the test section. For instance, in thermally induced microflows around heated Janus microparticles, the tracers are subject to thermophoretic forces that repel them from the heated part, thereby forming a tracer-depleted region. Such low-tracer-density regions (i.e., tracer-sparse regions), including diffusion-limited cases, are ill-suited for \mPIV~or \mPTV. Thus, controlling tracer positions at desired locations is helpful when the regions of interest are small and a large number of samples is required. In this paper, we propose the optical trapping of tracers as a solution.

Optical trapping was proposed by Ashkin \cite{Ashkin1970} to manipulate the motion of tiny particles using a focused laser. Practical guides for in-house experimental setups for optical trapping are available \cite{Neuman2004,Shaevitz2006}, and the technique has been widely applied to life sciences, nanosciences, and analytical chemistry. In flow measurement, optical trapping has been introduced in various forms \cite{Neve2008,Knoener2005,AlmendarezRangel2018,Nedev2014,Bruot2021,Koehler2016,Harlepp2017,Wei2019,Dehnavi2020,Kim2025,Zhao2025,Dara2023,Dara2025,DiLeonardo2006,Padgett2011a,Mushfique2008}. In Ref.~\cite{Neve2008}, the optical trapping of a target microparticle was used to investigate effectively the flow around the target by \mPIV. In Refs.~\cite{Knoener2005,AlmendarezRangel2018}, two methods using optical trapping were proposed. The first method (hereafter referred to as method 1) uses the balance between drag and optical forces, that is, the displacement of the optically-trapped tracers due to the flow is measured by camera or quadrant photodetectors. This method is called optical tweezer-based velocimetry (OTV). The OTV (or its variants) has been applied to measuring oscillatory microflows \cite{Nedev2014,Bruot2021}, rotlets \cite{Koehler2016}, in-vivo blood flows \cite{Harlepp2017}, the flow field around the cilia of microorganisms \cite{Wei2019,Dehnavi2020}, the measurement of three-dimensional flows \cite{Zhao2025}, and the Marangoni flows around thermoplasmonic bubbles \cite{Dara2023,Dara2025}. 
In Ref.~\cite{Kim2025}, many-body hydrodynamic interactions between colloidal particles exhibiting translation and rotation have been quantified using the optical trapping of partially fluorescent particles. 
The second method (say, method 2) is basically a standard PTV, but the measurement position is controlled by positioning the optically-trapped tracers as desired \cite{Knoener2005,AlmendarezRangel2018,DiLeonardo2006,Mushfique2008,Padgett2011a}. 
Although methods 1 and 2 have their own advantages, some challenges still remain. Method 1 (i.e., OTV) requires the calibration of the optical force to quantify the flow velocity. However, the optical force is sensitive to the fluid temperature and/or the diameter $d$ of the particle, since its magnitude scales as $\propto d^{n}$ $(n\geq3)$ \cite{Harada1996}. Moreover, because any misalignment of optical components or temporal changes in the entire apparatus can affect the magnitude of the optical forces, careful and persistent maintenance is necessary for reproducible experiments. Method 2 requires a large number of samples to average out the effects of noise. To address this difficulty, Refs.~\cite{DiLeonardo2006,Padgett2011a} introduced the repeated cycle of the trap and release of tracers.

\begin{figure}[bt]
    \centering
    \includegraphics[width=\figwidth\linewidth]{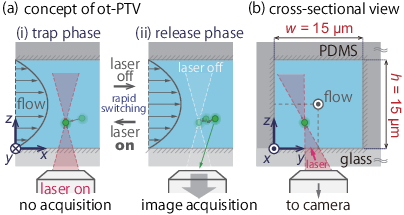}
    \caption{Schematic description of the present paper. (a) Concept of optically-trapped particle-tracking velocimetry (ot-PTV). In the trap phase, a tracer is captured at a measurement position using the optical force. In the release phase, the tracer follows the fluid flow in the absence of the optical force. These two phases are instantaneously switched and repeated. (b) Cross-sectional view of the microchannel. 
    The laser focus, i.e., the trap position of a tracer in the $y$ direction plane, can be controlled by AOD.}
    \label{fig:setup-main}
\end{figure}

In this paper, we propose optically-trapped particle tracking velocimetry (ot-PTV) as a solution to the above-mentioned tracer-sparse cases. The concept is simple, consisting of two stages, as illustrated in Fig.~\ref{fig:setup-main}(a). First, an optically-trapped tracer is held at a measurement position (a trap phase). Second, the tracer is released and advected in a flow (a release phase), during which sequential images are acquired. Then, after image acquisition, the same tracer is pulled back to the measurement position by the optical force, returning to the trap phase. By repeating the cycles of trap and release phases, we can achieve rapid image acquisition for the same measurement position. Note that the ot-PTV does not require quantification of the optical force, which was necessary in OTV. Although the concept of trap-and-release strategy was demonstrated in Ref.~\cite{DiLeonardo2006}, the detailed statistical analysis and the connection to the tracer-sparse cases have not been discussed. As a proof of concept, we present a validation of ot-PTV using pressure-driven microflows in a straight microchannel, with a particular interest in the analysis of slow flows with speeds on the order of $O(1)$~\um/s. The measurement of these slow flows is important, e.g., in thermally-induced microflows \cite{Tsuji2023}. As an application example of the proposed ot-PTV, preliminary results of the thermally-induced-microflows experiment are also given at the end of the paper in Sec.~\ref{sec:optothermal}.




\section{\label{sec:experiment-overview}Experiments}
Experimental details are given in \SI~\ref{sec:SI-experiment} \cite{Supp-THT2024}, and we provide an overview below. Note that only steady flows are considered here. 


\subsection{Setup}

The test section is a straight microchannel with a square cross-section. The width $w$ and height $h$ of the cross-section are both $15$~\um, as shown in Fig.~\ref{fig:setup-main}(b). 
An aqueous solution containing tracers fills the microchannel. Then, the fluid flow is driven by a pressure gradient $|\nabla_x p|$ with a resolution of approximately $0.01$~Pa. The $x$ component of the flow is denoted by $V=V(y,z)$. Using the incompressible Navier--Stokes equation and no-slip boundary condition on the channel walls, an analytical solution $V(y,z)$ is given as \cite{Bruus2007}
\begin{align}
&V(y,z)=\frac{4 h^2}{\pi^3 \mu} |\nabla_x p| \sum_{n,\,\text{odd}}^\infty 
\frac{1}{n^{3}}\left[1-\frac{\cosh(\frac{n\pi w}{2h}-\frac{n\pi y}{h})}{\cosh(\frac{n \pi w}{2h})}\right]\sin\left( \frac{n\pi z}{h}\right), \label{eq:flow} 
\end{align}
where $\mu$ is the fluid viscosity and $|\nabla_x p|$ is assumed constant.

The tracers are fluorescent, made of polystyrene (PS), and the tracer diameter is $d=1$~\um. The Stokes number of the tracer is much smaller than unity for a typical microfluidic setup (say, reference length $\approx 10$~\textmu m and reference speed $<1$~m/s; see \SI~\ref{sec:SI-sample-solution} \cite{Supp-THT2024}), and thus tracers follow the fluid flows. Moreover, the Reynolds number of the flow around the tracer is estimated as $Ud/\nu=10^{-5}\ll1$ with the reference flow speed $U=10$~\textmu m/s and the kinematic viscosity $\nu=1\times 10~^{-6}$~m$^2$/s, allowing us to consider the tracer in the Stokes-flow regime.

The tracers are optically trapped at a measurement position, as shown in Fig.~\ref{fig:setup-main}(b). The $y$ positions of the trapped tracer are controlled by electrical input signals to the acousto-optic deflector (AOD). The $z$ position of the trapped tracer is controlled by the $z$ position of the focal plane. The switching between the trap and release phases [Fig.~\ref{fig:setup-main}(a)] should be as instantaneous as possible; at least, the transition time must be much shorter than the frame intervals of image acquisition. Therefore, we use the AOD, which has a response time on the order of microseconds. To be more specific, we move the focal spot away during the release phase, rather than shutting out the laser using mechanical shutters. 

The power of the trapping laser is estimated to be $\approx24$~mW at the measurement position. 
According to Refs.~\cite{Ito2007,Peterman2003}, the temperature increase caused by the laser absorption is less than $1$~K with this power, due to the low absorption coefficient of water $14.2$~m$^{-1}$ and that of polystyrene $6$~m$^{-1}$ at the wavelength of $1064$~nm. The effects of thermo-osmosis and thermophoresis caused by this temperature increase are estimated as $<0.5$~\textmu m/s \cite{Cordero2009,Tsuji2021,Tsuji2023,Piazza2008} (see \SI~\ref{sec:compute-thermoosmosis-thermophoresis-effect} \cite{Supp-THT2024} for details).
Therefore, we neglect the effect of the laser-induced temperature increase.

The tracer diameter determines the spatial resolution of the flow measurement. Therefore, in principle, the smaller the tracer diameter, the higher the resolution. However, as the tracer diameter decreases, the optical gradient force that traps the tracer particles is significantly weakened \cite{Harada1996}, and, in addition, Brownian motion becomes more intense. These factors are the main bottlenecks of using smaller tracers. In our preliminary experiments with several tracer diameters, we find that a tracer diameter of $1$~\um~is practically suitable in our setup. 
For this tracer diameter, the radial range $r_c$ of the region in which we can pull back the tracers to the initial position is obtained as $r_c \approx 1$~\textmu m, using the estimate on the optical force for the Gaussian beam \cite{Ito2010} (see \SI~\ref{sec:optical-force} \cite{Supp-THT2024} for details). The critical range $r_c$ and the magnitude of the optical force both become smaller as the diameter $d$ decreases, thus leading to more frequent failure of optical trapping.


\subsection{\label{sec:procedure}Procedure}

\begin{figure}[bt]
    \centering
    \includegraphics[width=\figwidth\linewidth]{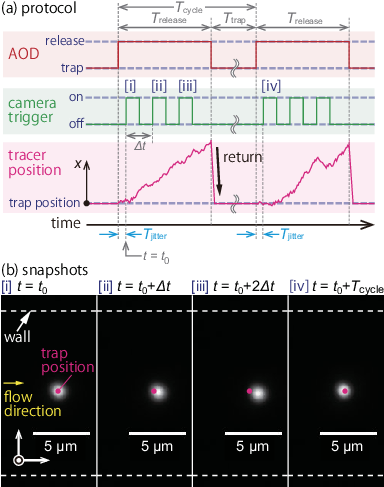}
    \caption{Schematic illustration of the experimental protocol. (a) AOD alternates between the trap and release phases. The time duration for the trap and release phases is $\Ttrap=40$~ms and $\Trelease=60$~ms, respectively. Trigger signals are sent to the camera to acquire images during the release phase. A time jitter between the start of the release phase and the first acquisition is $\Tjitter\approx 6$~ms. The released tracer is advected in the flow direction on average. Note that the tracer position curve is hand-drawn for schematic description. (b) Snapshots during a release phase ([i]--[iii]); A snapshot in the subsequent release phase [iv].}
    \label{fig:protocol}
\end{figure}

The experimental procedure is summarized in Fig.~\ref{fig:protocol}. First, the measurement position is determined using an optically-trapped tracer. To find the bottom wall (i.e., $z=0$~\um), we move the trapped tracer in the $z$ direction by adjusting the objective slowly, and look for a specific $z$ position at which the tracer starts to become defocused. We define this position as $z=0$~\um. Then, we move the tracer to the measurement position with the desired $(y,z)$ coordinates. In ot-PTV, we repeat the sets of the trap and release phases sequentially. As shown in Fig.~\ref{fig:protocol}(a), AOD alternates between the trap and release phases, which have the time durations of $\Ttrap=40$~ms and $\Trelease=60$~ms, respectively. The release time duration of $60$~ms is determined in preliminary experiments in accordance with the target flow speed, which is less than $10$~\textmu m/s in this paper. A longer duration would result in more frequent failure of trapping of tracers; a shorter release duration would result in smaller displacements and thus a lower signal-to-noise ratio. In the case of the failure of trapping, we discard the associated data in the present paper, to make the number of samples the same for all measurement positions. 
To reduce the amount of unnecessary data, images are acquired only during the release phases. 

Figure \ref{fig:protocol}(b) shows the snapshots of the tracer during a release phase ([i]--[iii] in Fig.~\ref{fig:protocol}) and the subsequent release phase ([iv] in Fig.~\ref{fig:protocol}) (see also \SI~\ref{sec:movie} \cite{Supp-THT2024} for movie). It can be seen from the snapshots [i] and [iv] that tracer positions in the first frames of the release phases are almost identical due to the return to the measurement position in the trap phase. This enables us to acquire the tracer displacements originating from the same measurement position. The error in the initial positions (see a slight difference between the positions of the tracer in snapshots [i] and [iv]) is due to the timing jitter $\Tjitter$ in Fig.~\ref{fig:protocol}(a). We define $\Delta x_0$ as the displacement of the tracer's initial position from its ensemble average. Then, we confirm that $\Delta x_0$ obeys the Gaussian distribution with the standard deviation of $123$~nm (see \SI~\ref{sec:initial} \cite{Supp-THT2024}), which represents the uncertainty of the measurement position. The timing jitter can be reduced further to decrease the uncertainty, as long as experimenters are sure that the acquisition duration does not include the trap phase. The acquired image sequence is then analyzed using the software ImageJ and its plug-in PTA (Particle Track and Analysis, version 1.2). 

Finally, some technical remarks are given as follows. (i) The fluorescence intensity decreases over time. Therefore, the trapped tracer is replaced by a fresh one occasionally in the sequential data acquisition. (ii) The optical trapping of single tracers is less feasible when the measurement position is located at a large $z$ value. More specifically, in the upper half region of the microchannel $z>7.5$~\um, the possibility of trapping two or more tracers increases, because the laser basically pushes upward (i.e., in the positive $z$ direction) the tracers passing through the region lower than the measurement position. Therefore, the test section is limited to the lower half region $z<7.5$~\um.


\section{Results and Discussion}\label{sec:result}

The set of measured positions is listed as follows: 
$y=3.4$, $4.5$, $5.5$, $6.6$, $7.6$~\textmu m and 
$z=1$, $2$, $3$, $5$, $6$~\textmu m. 
That is, $25(=5\times5)$ measurement positions are evaluated.
It should be noted that $y$ positions cannot be set closer to the wall because the optical trapping of the tracers becomes unstable near the wall. We consider that this instability is possibly due to the interference of the laser with the wall. 

\begin{figure}[bt]
    \centering
    \includegraphics[width=\figwidth\linewidth]{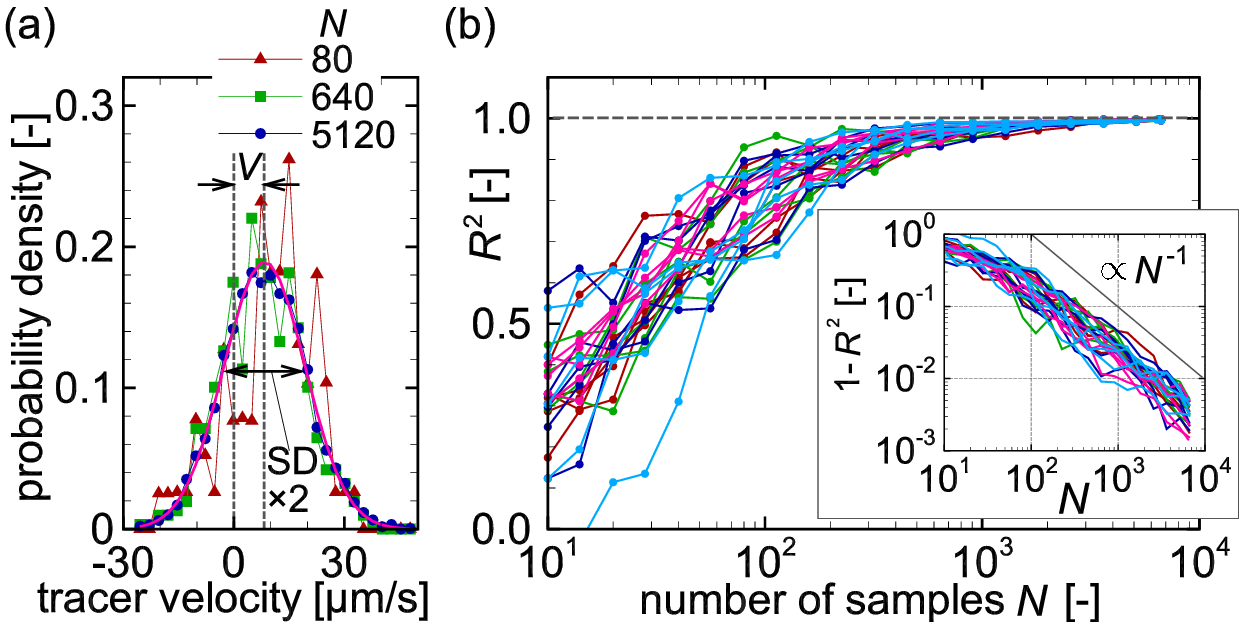}
    \caption{(a) Probability density of tracer velocity $v_x$ at $y=5.5$~\textmu m and $z=5$~\textmu m for sample sizes $N=80$ (red-triangle), $640$ (green-square), and $5120$ (blue-circle), where the bold-solid curve (magenta) is the Gaussian fit to the case of $N=5120$. In the panel, ``$V$" and ``SD" indicate the mean and the standard deviation, respectively. (b) Coefficient of determination $R^2$ for $y=3.4$~\textmu m (red), $4.5$~\textmu m (green), $5.5$~\textmu m (blue), $6.6$~\textmu m (magenta), and $7.6$~\textmu m (cyan), with $z=1$, $2$, $3$, $5$, and $6$~\textmu m. The inset shows $1-R^2$ in a double-logarithmic scale. }
    \label{fig:histgram-count}
\end{figure}

\subsection{Probability density of tracer velocities}

\subsubsection{Convergence of probability density to Gaussian}\label{sec:prob}

First, let us focus on the probability density of tracer velocities. 
Figure~\ref{fig:histgram-count}(a) shows a typical result for the probability density as a function of the tracer velocity $v_x$ at the measurement position $y=5.5$~\textmu m and $z=5$~\textmu m. Here, the average tracer velocity $\bvx$ is considered equivalent to the flow velocity $V$.
To construct the probability density, sample sizes of $N=80$, $640$, or $5120$ are used. 
The solid-magenta curve represents a Gaussian fit. 
Due to the central limit theorem, the probability density converges to the Gaussian distribution as the sample size increases. The average tracer velocity $\bvx$, i.e., the mean of the Gaussian distribution, is approximately $10$~\textmu m/s in Fig.~\ref{fig:histgram-count}(a), but this is not very clearly visible when the sample size $N$ is small (i.e., $N=80$). 
The standard deviation (SD) of the probability density [Fig.~\ref{fig:histgram-count}(a)] is given by SD$=\sqrt{2 D \Delta t}/\Delta t$, where $2 D \Delta t$ indicates the mean-squared-displacement (MSD), $\Delta t=12.5$~ms is the frame interval, and $D$ is the diffusion coefficient.
The data of SD at all measurement positions result in $D=0.63\pm0.03$~\textmu m$^2$/s. 
This value is well comparable to the diffusion coefficient computed as $D=0.55$~\textmu m$^2$/s using the Stokes-Einstein relation, where the diameter $d=1$~\textmu m, the viscosity $0.8\times10^{-3}$~Pa~s, and $T=300$~K are used as a representative case. Therefore, the randomness of the data stems from the diffusion of tracers.

Note that, if the flow speed is sufficiently high (e.g., 1~mm/s), the smallness of $N$ is not harmful since the mean is sufficiently away from zero, that is, the signal-to-noise ratio [or equivalently, mean-to-standard deviation ratio] is favorable. Therefore, the difficulty of small $N$ is specific to the measurement of slow microflows on the order of $O(1)$~\textmu m using small Brownian tracers. As Fig.~\ref{fig:histgram-count} demonstrates, the ot-PTV can offer reliable results even under poor signal-to-noise ratio (i.e., low $V/\mathrm{SD}$). Therefore, the lower bound of the measurement speed is not determined by the Brownian characteristics, but rather the image acquisition and image analysis processes (e.g., pixel size, the accuracy of particle position identification). 

Figure~\ref{fig:histgram-count}(b) shows the coefficient of determination $R^2$ (i.e., the goodness of fit) for various values of $y$ and $z$, where the inset displays the deviation from unity, i.e., $1-R^2$, in a double-logarithmic scale. The values of $R^2$ for all the cases converge to unity as the sample size $N$ increases, with the rate of convergence proportional to $N^{-1}$. 
Therefore, the probability densities at all the measured positions can be well approximated by the Gaussian distributions, indicating the adequacy of the obtained samples.  

Let us make a small remark on the advantage of ot-PTV. In ordinary PTV, one has to wait until tracers reach the region of interest (i.e., diffusion-limited). If the supply of tracers to the region of interest is not easy, obtaining a large number of samples using ordinary PTV would be problematic. Such a case can happen, e.g., (i) when the length scale of the region of interest is small due to the steep spatial variation of the flow, or (ii) tracers are scarce in the region of interest due to, e.g., external forces. The latter case will be demonstrated in Sec.~\ref{sec:optothermal} using the case of flows around a heated particle, where tracers tend to deplete from the region of interest due to thermophoretic forces.

\subsubsection{Position dependency of probability density}

\begin{figure}[bt]
    \centering
    \includegraphics[width=\figwidth\linewidth]{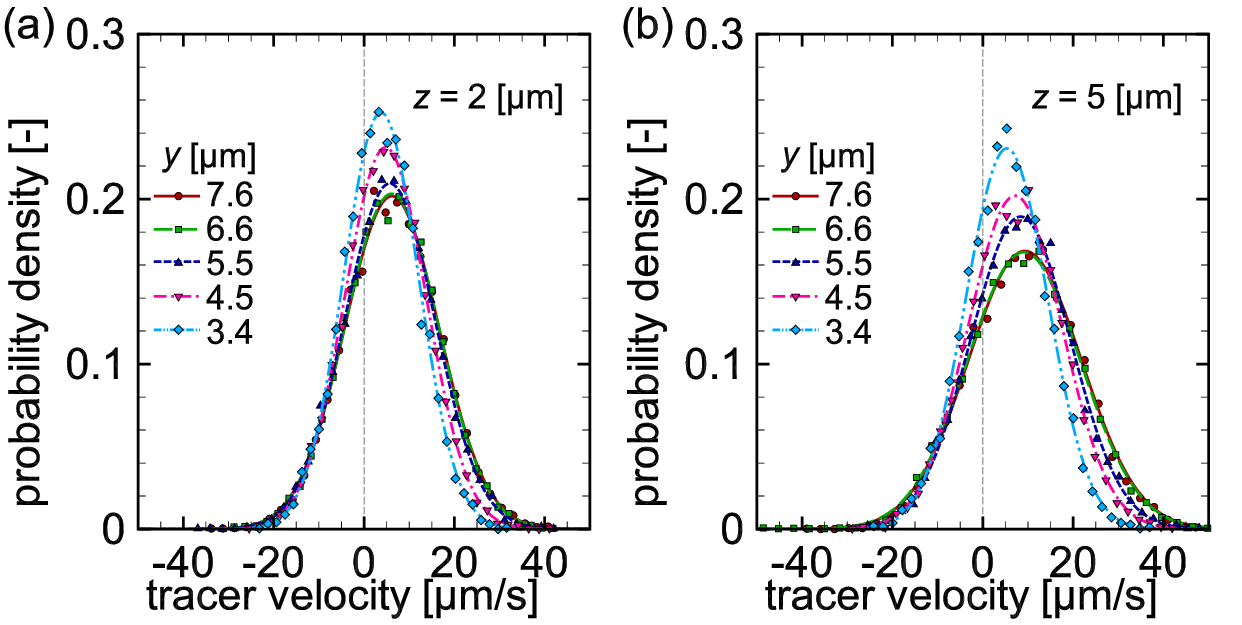}
    \caption{Probability density of tracer velocity $v_x$ at (a) $z=2$~\textmu m and (b) $5$~\textmu m for $y=7.6$~\textmu m (red-circle), $6.6$~\textmu m (green-square), $5.5$~\textmu m (blue-triangle), $4.5$~\textmu m (magenta-lower-triangle), and $3.4$~\textmu m (cyan-diamond). The curves show the corresponding Gaussian fits. The sample size is $N>6600$ for all cases. Results for other values of $z$ are presented in \SI~\ref{sec:SI-probability-density} \cite{Supp-THT2024}.}
    \label{fig:histgram-selected}
\end{figure}

Next, let us focus on the position-dependence of the probability densities. 
Figures~\ref{fig:histgram-selected}(a) and \ref{fig:histgram-selected}(b) show the probability densities at $z=2$~\textmu m and $z=5$~\textmu m, respectively, for various values of $y=3.4$, ..., $7.6$~\textmu m. All the experimental results (symbols) are well fitted by the Gaussian distributions (curves). As $y$ varies from $y=3.4$~\textmu m to $7.5$~\textmu m, the probability density varies as well, shifting the mean from smaller to larger magnitudes. The differences among the means are quite subtle. In other words, to distinguish flow velocity differences at different $y$ positions in the present flow conditions, the accurate reconstructions of the probability densities are required. This is the reason why we need a large number of samples for slow microflow measurements. 

It should be noted that the SD of the probability densities in Fig.~\ref{fig:histgram-selected} increases as $y$. A similar tendency has been reported in other microflow experiments such as Refs.~\cite{Jin2004,Huang2006}, where the finite-thickness imaging range was considered to be the cause. In our experiments, since ot-PTV analyzes the tracers at the same position, the cause is considered different. The SD contains the information about temperature, viscosity, and particle diameter, and thus the quantification of SD can lead to the measurement of these quantities. However, since only the mean values are relevant in velocimetry, we leave the origin of the varying SD as future work. 

\begin{figure}[bt]
    \centering
    \includegraphics[width=\figwidth\linewidth]{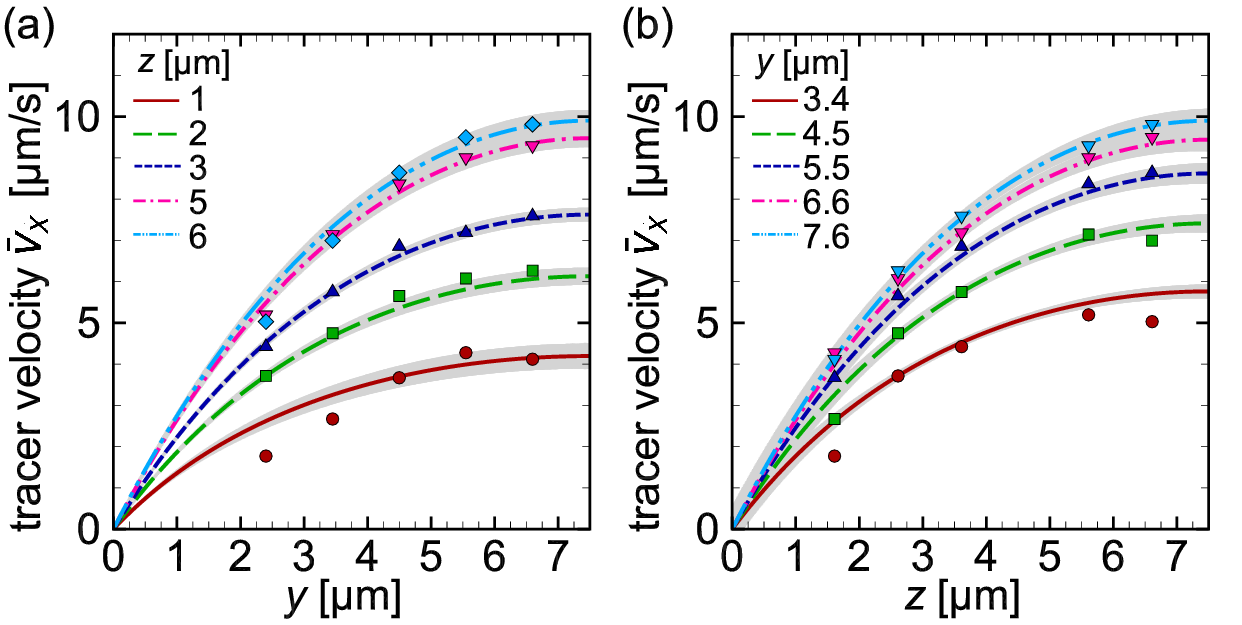}
    \caption{Average tracer velocity $\bvx$ as a function of (a) $y$ and (b) $z$, where symbols show the experimental data, curves indicate the results of fitting using Eq.~\eqref{eq:flow}, and shaded regions indicate the 95\% confidence interval. Note that experimental data (symbols) are shifted by (a) $\delta y=-1.01$~\textmu m in the $y$ direction and (b) $\delta z=0.61$~\textmu m in the $z$ direction to consider sub-micron scale errors due to the difficulty of microscopic observation. (a) $z=1$~\textmu m (red-circle), $2$~\textmu m (green-square), $3$~\textmu m (blue-triangle), $5$~\textmu m (magenta-lower-triangle), and $6$~\textmu m (cyan-diamond), and (b) $y=3.4$~\textmu m (red-circle), $4.4$~\textmu m (green-square), $5.5$~\textmu m (blue-triangle), $6.5$~\textmu m (magenta-lower-triangle), and $7.6$~\textmu m (cyan-diamond).}
    \label{fig:profile}
\end{figure}

\subsection{Flow velocity analysis}

\subsubsection{Spatial profiles}

Figure~\ref{fig:profile} summarizes the measurement results of $\bvx(=V)$ at 25 positions. 
Figures~\ref{fig:profile}(a) and \ref{fig:profile}(b) show the same data plotted as a function of $y$ and $z$, respectively. 
The experimental results (symbols) are fitted by the analytical solution (curves), Eq.~\eqref{eq:flow}.

First, let us explain the fitting method. 
In Eq.~\eqref{eq:flow}, we assume that the viscosity $\mu=1\times 10^{-3}$~Pa$\cdot$s and the channel dimensions $h=w=15$~\textmu m are given.
Therefore, $|\nabla_x p|$, $y$, and $z$ are treated as fitting variables. 

The pressure gradient is recast as $|\nabla_x p|=\Delta p/L_x$ where $\Delta p$ is the pressure difference across the microchannel of the length $L_x=500$~\textmu m (see \SI~\ref{sec:SI-microfluidic-device} and \ref{sec:SI-pressure-gradient} \cite{Supp-THT2024}). 
The coordinates at measurement positions are specified as $y=3.4$, ..., $7.6$~\textmu m and $z=1$, ..., $6$~\textmu m in the experiments, using the method described in Sec.~\ref{sec:procedure}. However, it is possible that these coordinates contain systematic sub-micron scale errors due to the difficulty of microscopic observation. 
Therefore, we replace $(y,\,z)$ of Eq.~\eqref{eq:flow} by $(y+\delta y,\,z+\delta z)$ in the fitting process, and search for a set of $(\Delta p,\,\delta y,\,\delta z)$ that minimizes the mean squared error with respect to the experimental data. 
The values $(\Delta p,\,\delta y,\,\delta z)$ for the best fit are found to be $\Delta p=0.302$~Pa, $\delta y=-1.01$~\textmu m, and $\delta z=0.61$~\textmu m. We consider that the pressure difference $\Delta p = 0.302$~Pa estimated from the fitting to be reasonable, since the pressure difference estimated from the experimental condition is $\Delta p= 0.294\pm 0.039$~Pa, as described in \SI~\ref{sec:SI-pressure-gradient} \cite{Supp-THT2024}. The values of $\delta y$ and $\delta z$ are also reasonable considering the observational and fabrication accuracies of the present experimental system.

Next, let us focus on the results of the fitting in Fig.~\ref{fig:profile}. 
It is evident that the experimental values are well fitted by the analytical profile in Eq.~\eqref{eq:flow}. 
We remark that the measured flow speeds ($<10$~\textmu m/s) are slow compared with the maximum instantaneous tracer speed ($>30$--$40$~\textmu m/s; see the probability density in Fig.~\ref{fig:histgram-selected}), and thus averaging over many samples is essential to obtain the accurate position dependence of the flow speed with a resolution of $O(1)$~\textmu m/s. 
It should be mentioned that the data at $z=4$ and $7$~\textmu m, which are not presented in Fig.~\ref{fig:profile}, are discarded from the figure by the observation of the probability distribution, as described in \SI~\ref{sec:SI-probability-density} \cite{Supp-THT2024}.

Finally, we make a comment on the comparison with a standard \mPIV. For the present simple flow with $x$ uniformity, only a $10$~s-measurement is enough to reproduce the theoretical expectation Eq.~\eqref{eq:flow} (see \SI~\ref{sec:piv} \cite{Supp-THT2024} for the detail of PIV analysis). However, there are some cases where the \mPIV~will not be effective, e.g., (i) the spatial averaging is not available; (ii) the tracers are scarce in the region of interest due to some external forces. As for (i), if we do not take the spatial average in the $x$ direction for this PIV analysis, a longer operation with several minutes will be necessary to reduce the statistical noise. The latter case (ii) will be discussed later in Sec.~\ref{sec:optothermal}.

\subsubsection{Effect of the number of samples}\label{sec:effect-of-number}
\begin{figure}[bt]
    \centering
    \includegraphics[width=\textwidth]{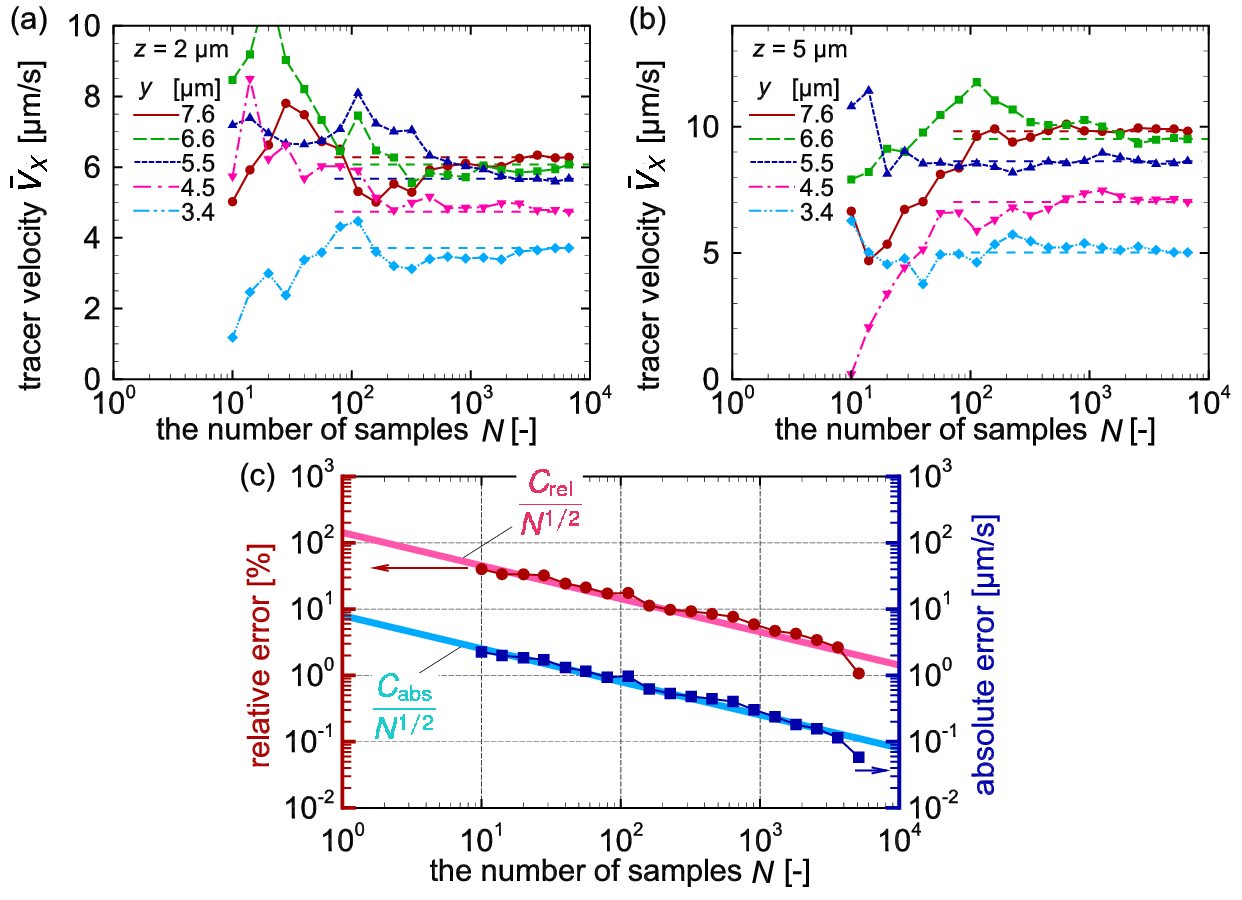}
    \caption{Average tracer velocity $\bvx$ at (a) $z=2$~\textmu m and (b) $z=5$~\textmu m as a function of the number of samples $N$ for various $y$. See also Fig.~\ref{fig:histgram-selected} for the corresponding probability density distributions at the same position $(y,\,z)$. (c) Relative error (left axis, red circle) and absolute error (right axis, blue square) as a function of $N$ averaged over whole measurement positions. The dashed lines for relative and absolute errors represent $\propto N^{-1/2}$ with proportionality constants $C_{\mathrm{rel}}=160$~\% and $C_{\mathrm{abs}}=8$~\textmu m/s, respectively.}
    \label{fig:vx-vs-N}
\end{figure}

To reduce the experimental duration, it is useful to know the optimal number of samples $N$ which is required to realize the desired accuracy. Figures~\ref{fig:vx-vs-N}(a) and \ref{fig:vx-vs-N}(b) show the average tracer velocity $\bvx$ as a function of $N$ at $z=2$~\textmu m and $z=5$~\textmu m for various $y$, respectively (see also Fig.~\ref{fig:histgram-selected} for the corresponding probability densities). In the figure, the horizontal dashed lines represent the results of full samples with $N>6000$. It is vaguely seen that $\bvx$ approaches a certain value as $N$ increases, as expected, but the manner of approach is not clear due to statistical noise. Qualitatively, the number of samples in the range $100<N<1000$ seems to yield reasonable results under the present experimental condition with $\bvx<10$~\textmu m/s. 

For more quantitative evaluation, Fig.~\ref{fig:vx-vs-N}(c) shows the relative and absolute errors of $\bvx$, which are calculated in the following manner. The relative error, shown by red circle and left axis, is the average over all 25 measurement positions of $|\bvx^{\mathrm{(ref)}}-\bvx^{(N)}|/\bvx^{\mathrm{(ref)}}$, where $\bvx^{\mathrm{(ref)}}$ is a reference value obtained with whole samples $(N>6000)$ and $\bvx^{(N)}$ is the value obtained with first $N$ samples. Therefore, to obtain the results with a relative error less than $10$~\% for the present flow speed condition ($<10$~\textmu m/s), the figure indicates that $N$ should be $N>200$. The absolute error, shown by the blue square and right axis, is the average of $|\bvx^{\mathrm{(ref)}}-\bvx^{(N)}|$ over the whole measurement positions. To obtain the results with absolute error less than $1$~\textmu m/s, $N$ is required to be $N>200$.

The bold straight lines in Fig.~\ref{fig:vx-vs-N}(c) represent the best fit using power law functions $\propto N^{-1/2}$. For the relative error (red circle), the power law function is $C_{\mathrm{rel}}/N^{1/2}$ with $C_{\mathrm{rel}}=143$~\%. Note that $C_{\mathrm{rel}}$ indicates the relative error when only a single sample ($N=1$) is used for the analysis. For much higher flow speed, $C_{\mathrm{rel}}$ is expected to decrease since only the denominator increases while the numerator is determined by the randomness of the data characterized by the Brownian motion.
In the same manner, for the absolute error (blue square), the power law function is $C_{\mathrm{abs}}/N^{1/2}$ with $C_{\mathrm{abs}}=8.0$~\textmu m/s. This value of $C_{\mathrm{abs}}$ is well consistent with the SD of the probability density, shown in Figs.~\ref{fig:histgram-count} and \ref{fig:histgram-selected}. In fact, using the data of the whole measurement position, the SD of the probability density is computed as $10.0\pm0.2$~\textmu m/s. In this way, we can evaluate the expected error in the ot-PTV.

The drawback of the ot-PTV is that it is a pointwise method, and thus a long experimental duration with area scanning is required when two-dimensional profiles are under concern. To address this difficulty, reducing experimental duration is possible by optimizing the number of samples $N$, as demonstrated above. In our case, the number of samples may be reduced to $N\approx 200$, which results in a $30$-fold speed up, sacrificing the accuracy of $1$~\textmu m/s.

\subsection{Features of ot-PTV}\label{sec:make-comparison}
We discuss some advantages and disadvantages of ot-PTV in this section, by comparing it with other methods. 

\subsubsection{Comparison with \mPIV~and TIRV}
As mentioned in Sec.~\ref{sec:intro}, flow measurements near boundaries are one of the important issues in the microfluidics community. 
This is mainly due to velocity slips, which have been investigated experimentally using \mPIV~\cite{Tretheway2002,Joseph2005}. The effects of the hydrophilic surface properties on slip lengths were pointed out in these references. 
To further improve spatial resolution near boundaries, the use of evanescent waves as an excitation light source was introduced \cite{Zettner2003,Jin2004}. 
This method is called total internal reflection velocimetry (TIRV), and has been applied to slip-length measurements \cite{Huang2006,Huang2007}. The use of evanescent waves was further applied to flow measurements in nanochannels \cite{Kazoe2013}. With the aid of an optical image analysis technique using defocusing, the measurement of nanoscale flow profiles was further refined \cite{Kazoe2021}. 

The main differences between the present ot-PTV and the above-mentioned preceding studies are found in the range of the flow speed under investigation. The measured flow speed ranges were, approximately, 
$2$--$40$~\textmu m/s in Ref.~\cite{Zettner2003}, 
$30$--$300$~\textmu m/s in Ref.~\cite{Jin2004}, 
$50$--$500$~\textmu m/s in Ref.~\cite{Hu2006}, 
$>1$~mm/s in Refs.~\cite{Kazoe2013,Kazoe2021}. Although Ref.~\cite{Zettner2003} measured the flows with relatively slow speeds, the measured velocities were the spatial average over a certain distance near the boundaries. 
In contrast, the present ot-PTV focuses on the flow speed $<10$~\textmu m/s and resolves the spatial profiles as functions of the distances from the boundaries. 
It should also be noted that TIRV is applicable to flows over a planar boundary, but ot-PTV can be used, in principle, for flows over curved surfaces such as those around spheres \cite{Tsuji2023}. 

\subsubsection{Comparison with MTV and LIFPA}
Another class of velocimetry techniques are molecular tagging velocimetry (MTV) \cite{Hu2006,McKeon2007} and laser-induced fluorescence photobleaching anemometer (LIFPA) \cite{Wang2005a,Kuang2009,Kuang2010,Kuang2011}. 
In these methods, molecules are added to a working fluid and are tagged by lasers via photochemical reaction such as phosphorescence and photobleaching. The subsequent motion of tagged molecules is tracked by imaging and analyzed to obtain flow velocities. 
While MTV has been applied to the measurement of flow spatial profiles, LIFPA is specialized for pointwise measurements with a high temporal resolution. For microscope analysis in microfluidics and nanofluidics, LIFPA is considered more convenient, since a single objective lens can be used for both laser irradiation and observation \cite{Kuang2009,Kuang2010,Kuang2011}. 

The main difference between the ot-PTV and the LIFPA is the range of the flow speed under investigation. The LIFPA utilizes molecular motion, which consists of the combination of both translational and diffusive components. Due to the strong diffusion of the molecules, the flow speed that can be measured by LIFPA is relatively high: for instance, Refs.~\cite{Kuang2009,Kuang2010} mainly considered the range $> 1$~mm/s. 
In contrast to LIFPA, the ot-PTV can treat much slower flows with the resolution of $1$~\textmu m/s, as already discussed above in this section. To obtain flow profiles, both LIFPA and ot-PTV require scanning across a test section, which is a drawback compared to \mPIV, \mPTV, TIRV, and MTV.

\subsubsection{Comparison with OTV}
As introduced in Sec.~\ref{sec:intro}, also OTV uses optical trapping of tracers, although the measurement principle is totally different from that of ot-PTV. In Refs.~\cite{Dehnavi2020,Harlepp2017}, photo detectors are used to acquire tracer positions (supplemented by camera observation). Therefore, the sampling rate is much higher than the present ot-PTV analysis (say, several kHz \cite{Harlepp2017}). This high data acquisition rate enables the analysis of time-dependent periodic motion with $O(10)$~Hz with amplitude $1.5$--$70$~\textmu m/s \cite{Dehnavi2020}. Furthermore, the OTV is less affected by the Brownian motion, since the tracer is optically trapped during data acquisition. However, the experimental setup of OTV is much more involved than that ot-PTV here, requiring the optical-force calibration, the simultaneous use of both photo detector and camera, and data analysis using a Kalman filter. The ot-PTV setup is easier to introduce because calibration is not necessary and the standard PTV protocol can be used.

One possibility to improve ot-PTV for faster flow speed is to use a photo detector as in Refs.~\cite{Dehnavi2020,Harlepp2017} instead of using a camera. To realize this, we need to develop an algorithm to deduce particle velocity from a time-series data of intensity. This may be feasible, especially using fluorescent tracers, since they have a well-characterized intensity distribution.

\subsubsection{Some technical issues of ot-PTV}

Finally, the limitations and some technical issues of ot-PTV are explained. 
The durations of trap and release ($T_{\mathrm{trap}}$ and $T_{\mathrm{release}}$ in Fig.~\ref{fig:protocol}) are closely related to the expected flow speed. For instance, if $T_{\mathrm{release}}$ is too long, the optical trap cannot pull back the tracer to the measurement position; if $T_{\mathrm{release}}$ is too short, the obtained samples are more affected by the precision of particle-tracking image analysis. Since the optical force depends on available optical tools and the tracer diameter, trial and error will be necessary to find a suitable protocol of trap and release in practice. 
Another issue is the total time duration of experiments. In this paper, a single run to measure a point takes $330$ s for full sample numbers $N=6600$ (i.e., $2$ samples per cycle and $10$ cycles per second). Therefore, to obtain a profile (e.g., $30$ measurement points in a test section), it will take at least $3$ hours. Such a long duration may cause undesired drifts in experimental parameters (e.g., flow speeds, observation position, adsorption of tracers, temperature, etc.), and should be avoided. To address this issue, the use of multiple optical traps using liquid-crystal-on-silicon devices will be useful; in Ref.~\cite{DiLeonardo2006}, a holographic optical tweezer using a spatial light modulator was applied for multiple optical traps. By choosing an appropriate size of $N$ in accordance with the requirement of the flow analysis (e.g., expected flow speed, allowable level of error; see Sec.~\ref{sec:effect-of-number}), together with the multiple optical trap strategy, we can reduce the experimental time significantly.


\begin{figure}[bt]
    \centering
    \includegraphics[width=\figwidth\linewidth]{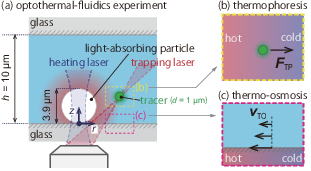}
    \caption{Schematic of the experimental system demonstrated as an application of ot-PTV. (a) Configuration of the system. Heating laser ($488$~nm wavelength) is focused on a light-absorbing particle ($d=3.9$~\textmu m); trapping laser ($1064$~nm) is focused on a fluorescent tracer ($d=1$~\textmu m). (b) Thermophoretic force $\bm{F}_{\TP}$ acts on the tracer in the opposite direction of the temperature gradient. (c) Thermo-osmosis $\bm{v}_\TO$ occurs along the surface in the direction of the temperature gradient. Thermophoresis and thermo-osmosis coincide upon laser heating of a light-absorbing particle [panel (a)].}
    \label{fig:optothermal}
\end{figure}

\section{Application}\label{sec:optothermal}

A more compelling test case, in which the advantage of ot-PTV is significant, is demonstrated here. The data presented here are preliminary, and more detailed research is ongoing and will be reported in future work.

\subsection{Experimental setup}

The experimental system is so-called optothermal fluidics, as shown in Fig.~\ref{fig:optothermal}(a). In this system, in addition to the laser with wavelength $1064$~nm for optical trapping, a visible focused laser with wavelength $488$~nm ($20$~mW) is irradiated to a gap of $h=10$~\textmu m between two glass substrates. In particular, this visible laser is focused onto a light-absorbing particle (diameter$=3.9$~\textmu m; PS-MAG-S2180, microParticle GmbH), which heats up upon laser irradiation (see Ref.~\cite{Paul2022a} for details; similar particles with a smaller diameter were used there). Therefore, the visible laser is referred to as the heating laser hereafter.

Due to the laser heating, we expect two dominant thermal effects on the fluid: (i) thermophoresis of tracers [Fig.~\ref{fig:optothermal}(b)] and (ii) thermo-osmosis along the substrate [Fig.~\ref{fig:optothermal}(c)]. In panel (b), $\bm{F}_{\TP}$ is a thermophoretic force that is proportional to the temperature gradient of the fluid (see, e.g., Refs.~\cite{Piazza2008,JimenezAmaya2025}). The force is directed toward the colder region in our test here, where the tris-HCl buffer of 1~mM is used as a solvent, as verified in Ref.~\cite{Tsuji2018a}. In panel (c), $\bm{v}_{\TO}$ is the thermo-osmotic flow velocity that is proportional to the temperature gradient of the substrate; thermo-osmosis is considered localized near the solid surface and directed toward the hotter region \cite{Bregulla2016,Fraenzl2022,Tsuji2023}.
Note that thermal convection is negligibly small because of the small gap size \cite{Paul2022a,Tsuji2024}, and the laser power is kept low enough so that the creation of bubbles never happens. The goal here is to investigate the flow characteristics of thermo-osmosis.

\subsection{Why ot-PTV? ---Some remarks---}\label{sec:why}

The advantage of ot-PTV over a standard \mPIV~or \mPTV~is described as follows. In this experiment, tracers are subject to a thermophoretic force directed toward the colder region. Therefore, even though the flow field (i.e., thermo-osmosis) is steady, the tracer distribution is unsteady and nonuniform, being scarce near the light-absorbing particle. These features make a standard \mPIV~inefficient. After a single experimental run, one must wait for the tracers to diffuse over the region of interest before carrying out the next run; in other words, the system is diffusion-limited. Since the expected flow speed is slow $<10$~\textmu m/s, many repeated runs are necessary to reduce statistical noise, but these diffusion-limited characteristics hinder efficient data acquisition.

\begin{figure}[bt]
    \centering
    \includegraphics[width=\textwidth]{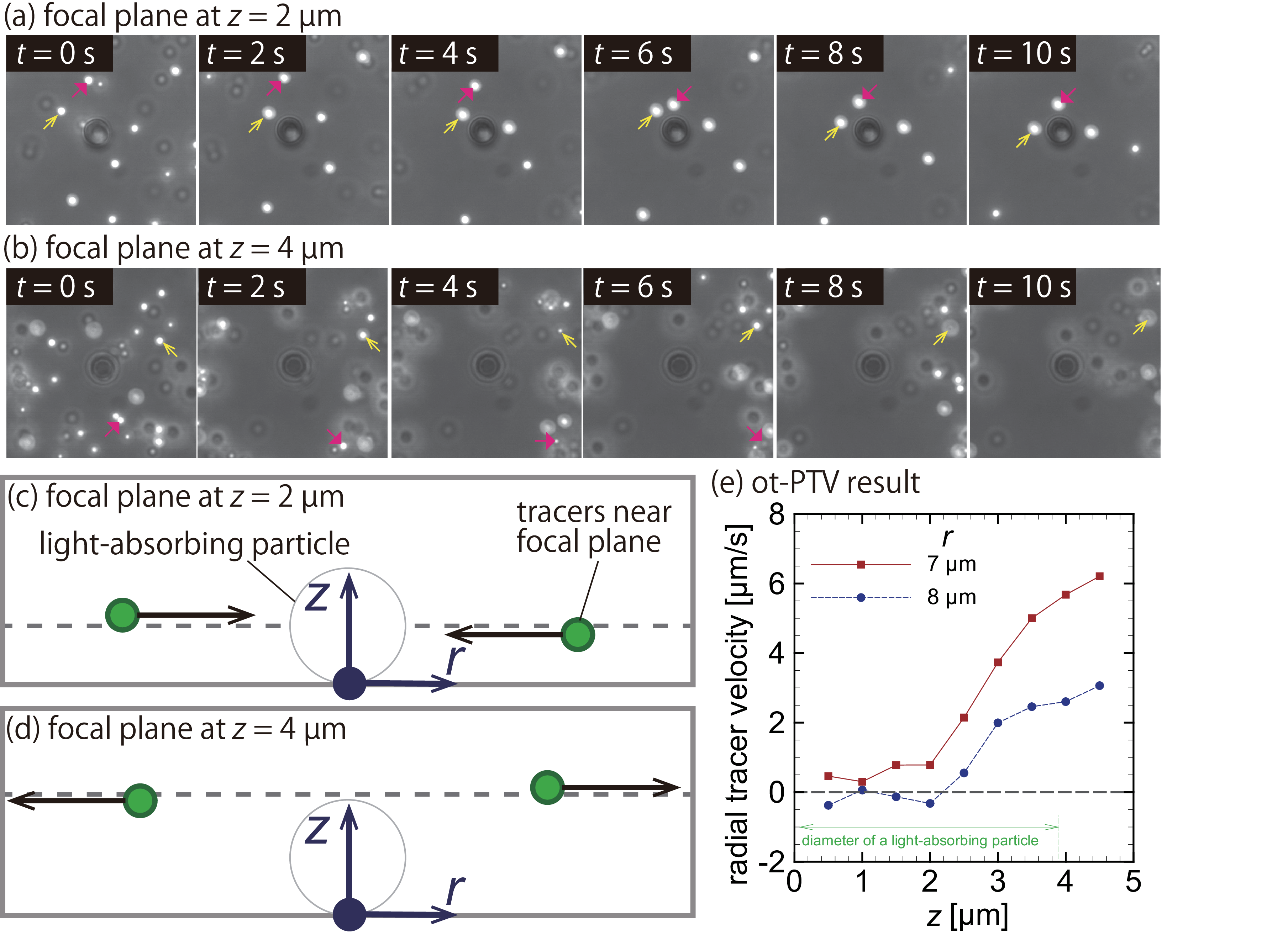}
    \caption{Snapshots of optothermal fluidics experiment with focal plane at (a) $z=2$~\textmu m and (b) $z=4$~\textmu m. Arrows indicate the same particles. See also movies in \SI~\ref{sec:movie} \cite{Supp-THT2024}. Panels (c) and (d) are rough sketches to describe the situations in panels (a) and (b), respectively. (e) Results of ot-PTV measurement. Radial component of tracer velocity is shown as a function of $z$ for $r=7$~\textmu m (red square) and $r=8$~\textmu m (blue circle).}
    \label{fig:application-results}
\end{figure}

To evaluate thermo-osmosis, we need to subtract the effect of thermophoresis of tracers. This needs two more steps: (i) estimate of the thermophoretic force and (ii) temperature profile measurement (or analysis using computational models). However, these are not directly related to the scope of the present paper, we leave these steps for future work, and evaluate only the tracer motion in the following.

\subsection{Results}
First, let us see the snapshots of the observed phenomena. Figure~\ref{fig:application-results}(a) and (b) show the cases of the focal plane set at $z=2$~\textmu m and $4$~\textmu m, respectively. See also movies in \SI~\ref{sec:movie} \cite{Supp-THT2024}. Time $t=0$~s indicates the start of the heating laser irradiation. Panels (c) and (d) are rough sketches to describe the motion of tracers in the snapshots (a) and (b), respectively. 

In panel (a), the tracers near the focal plane look like white circles, whereas defocused tracers look dark. The snapshots indicate that only tracers near the focal plane come close to the light-absorbing particle, keeping certain distances at $t=10$~s. However, the defocused tracers are repelled from the center, indicating the thermophoretic depletion. Conversely, in panel (b), the tracers near the focal plane are depleted from the heat. These situations suggest that (i) the tracers near the bottom surface are brought to the heat source by thermo-osmosis (toward hot) near the boundary, finding the $r$ position at which the effects of thermophoresis and thermo-osmosis are balanced [panel (c)]; (ii) the tracers away from the bottom surface are subject to only thermophoresis and are repelled from the heat [panel (d)].

As described in Sec.~\ref{sec:why}, a standard PIV approach is not efficient due to the tracer depletion. Therefore, to investigate the flow characteristics, ot-PTV may be helpful. Figure~\ref{fig:application-results}(e) shows the results of ot-PTV at $r=7$~\textmu m and $r=8$~\textmu m as a function of $z$. The number of samples is larger than $3000$ for all the data points. For $z\gtrsim3$~\textmu m, radial tracer velocity $v_r$ is positive and away from zero, indicating that the tracers are repelled by heat due to thermophoresis. The plots for $r=8$~\textmu m show a smaller magnitude because the temperature gradient decreases as $r$ becomes larger. Conversely, for $z\lesssim2$~\textmu m, $v_r$ stays near zero, and $v_r$ at $r=8$~\textmu m becomes negative (i.e., tracers move toward hot). However, the magnitude of $v_r$ for $z\lesssim2$~\textmu m is so small and within the statistical noise. We consider that the tracers near the bottom boundary only show ``Brownian motion," possibly under the counterbalance between thermophoresis and thermo-osmosis. These data are consistent with the observation of snapshots. Therefore, ot-PTV is expected to serve as a velocimetry to quantify the flow characteristics of optothermal fluidics.

\section{Conclusion}\label{sec:conclusion}
In this paper, we proposed optically-trapped particle tracking velocimetry (ot-PTV) and demonstrated the measurement of a pressure-driven flow in a microchannel with a square cross-section. By rapidly repeating the cycles of the trap and release phases of an individual tracer, several thousand samples (i.e., tracer displacements) were accumulated within minutes. Thanks to the large sample size, the probability densities of the tracer velocities sufficiently converged to Gaussian distributions, and the spatial profile of slow microflows $<10$~\textmu m/s was successfully measured in good agreement with a theoretical prediction. Despite the drawback that ot-PTV is a pointwise method having a lower data acquisition rate as compared to \mPIV, ot-PTV enables a well-controlled tracer seeding, which is advantageous in a situation where tracers are scarce in the test section.
The proposed ot-PTV was applied to the measurement of thermally-induced microflows, where the tracers are depleted from the test section due to thermophoresis. Although the results are preliminary, it was shown that the ot-PTV can be an efficient measurement technique even when tracers are scarce, a situation that is inefficient for \mPIV.

\begin{acknowledgments}
This work was supported by the Japan Society for the Promotion of Science KAKENHI Grants No. 24K00803 and No.~25K01156, and also by the Japan Science and Technology Agency PRESTO Grant No. JPMJPR22O7. The authors thank Mr. Shota Suzuki for the assistance in the experiment of Sec.~\ref{sec:optothermal}.
\end{acknowledgments}

\clearpage

\appendix
\onecolumngrid

\setcounter{page}{1}
\renewcommand{\appendixname}{\SI}

\begin{center}
\SI~on \\[1em]
{\large
{\bf \titleB}}
\\[1em]
{
Tetsuro Tsuji$^\ast$, Shoma Hashimoto, and Satoshi Taguchi\\[0.5em]
Graduate School of Informatics, Kyoto University, Kyoto 606-8501, Japan\\[0.5em]
$^\ast$tsuji.tetsuro.7x@kyoto-u.ac.jp
}
\end{center}


\setcounter{figure}{0}
\renewcommand\thefigure{S\arabic{figure}}   
\setcounter{table}{0}
\renewcommand\thetable{S\arabic{table}} 
\renewcommand\theequation{\thesection.\arabic{equation}} 
\renewcommand{\thesection}{S\arabic{section}}
\renewcommand{\thesubsection}{\Alph{subsection}}
\renewcommand{\thesubsubsection}{\Roman{subsubsection}}


\section{Experimental details}\label{sec:SI-experiment}

\subsection{Sample solution}\label{sec:SI-sample-solution}
We use fluorescent polystyrene (PS) particles with a diameter of $1$~\textmu m (F8823, Thermo Fisher Scientific) as flow tracers. We prepare an aqueous solution of the PS particles dispersion, where the non-ionic surfactant Triton X-100 (Sigma Aldrich) is also added to reduce the adsorption of the tracers to the channel walls. The concentrations of the PS particles and the surfactant are $1\times10^{-4}$~wt\% and $1$~wt\%, respectively. To avoid the optical trapping of multiple particles at the same time, the tracer concentration is relatively lower compared with usual particle tracking velocimetry (PTV). 

The relaxation time of the tracer is given as $\tau=\rho_p d^2/(18\mu)$, where $\rho_p \approx 1\times 10^3$~kg/m$^3$ is the mass density of PS, $d=1\times10^{-6}$ m is the diameter of the tracer, $\mu\approx 1\times 10^{-3}$~Pa$\cdot$s is the viscosity of water at room temperature. Therefore, $\tau$ is estimated as $\tau\approx 50$~ns, that is, the tracer velocity after release immediately approaches the terminal velocity, which is equal to the flow velocity. The data acquisition is made only in the release phase, and thus the acquired data never include the transient behavior toward the terminal motion. 

The Stokes number for reference speed $U$ and reference length $L$ is defined as $\mathrm{St}=\tau U/L$. If $\mathrm{St}\ll1$, we can assume that tracer motion well reproduces streamlines. For instance, let us take $L=10$~\textmu m for microflow observation. Then, we get $\mathrm{St}\approx (5\times 10^{-3}$~s/m$)\times U$, meaning that $U<1$~m/s leads to $\mathrm{St}=5\times 10^{-3}\ll1$. Thus, the tracer with $d=1$~\textmu m ensures the low-$\mathrm{St}$ limit in usual microfluidic experiments.

\subsection{Optical setup}\label{sec:SI-optical-setup}
\begin{figure}[b]
    \centering
    \includegraphics[width=\figwidth\linewidth]{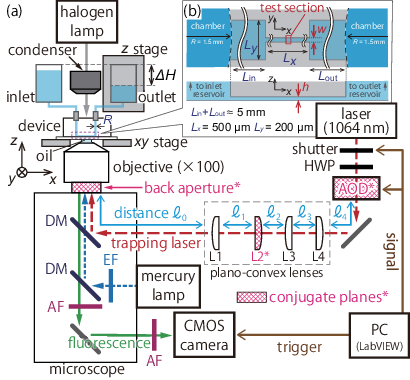}
    \caption{(a) Overview of the experimental setup. DM: dichroic mirror, AF: absorption filter, EF: emission filter, AOD: acousto-optic deflector, HWP: half-wave plate. (b) Detail of the microchannel.}
    \label{fig:SI-setup}
\end{figure}

Figure~\ref{fig:SI-setup}(a) shows an in-house optical system of the present paper. 
A laser with a wavelength of 1064~nm (Powerwave 1064, NPI laser) is used for optical trapping. The laser is guided to an inverted microscope (IX-73, Olympus), and is irradiated to a microfluidic device through the oil-immersion objective (UPLXAPO100XO, Olympus, $\text{numerical aperture}=1.45$).
The laser goes through the acousto-optic deflector (AOD) and plano-convex lenses (L1, ..., L4) before entering the microscope, where the back aperture of the objective, the lens L2, and the AOD are placed at conjugate planes \cite{Neuman2004,Shaevitz2006}. 
By placing AOD at the conjugate plane of the back aperture, we can manipulate precisely the laser focal position in the $xy$ plane of the test section without power loss. More specifically, by applying an electronic signal to AOD via the controller (PC with LabVIEW; PXIe-6363, PXIe-8861, PXIe-1071, NI), we can move the trap position (i.e., the laser focus) instantaneously. Note that, to place AOD at the conjugate plane, the distances $\ell_0$, ..., $\ell_4$ shown in the figure are determined according to the f values of the lenses. Technical recipes for the construction are given in Ref.~\cite{Shaevitz2006}. 

The response time of AOD is 1.58~\textmu s according to the specification sheet. Therefore, the time delay of AOD is negligible in our experiments, where frame intervals are of the order of milliseconds. The resolution of positioning the focal position in the $xy$ plane is estimated as $1.6$~nm for the present optical and electronic devices. We use an $xy$ auto-stage to locate the test section under the microscope view, and the measurement position (i.e., the focal position) in the test section is adjusted by the electrical input to AOD.


\subsection{Microfluidic device}\label{sec:SI-microfluidic-device}
Figure~\ref{fig:SI-setup}(b) shows the whole view of the microfluidic device made of polydimethylsiloxane (PDMS) and a glass substrate. 
This is a standard microfluidic device made of PDMS, but some details are given as follows. 

First, let us describe the fabrication process of the PDMS block. 
A silicon-on-insulator wafer with a device layer of thickness $15\pm 1$ \um~is purchased from Seiren KST Corporation. After fabricating on the wafer the patterned-photoresist layer of the channel pattern [Fig.~\ref{fig:SI-setup}(b)] using the photolithography technique, the bare part of the device layer is dry-etched using deep reactive-ion etching. Finally, we remove the photoresist by oxygen ashing. The master mold thus fabricated is used to make the PDMS block as follows. A 10:1 mixture of uncured PDMS and curing agent is poured in a chamber with the mold fixed to the bottom. The PDMS is then cured by heating. The cured PDMS block with the pattern of the microchannel at the bottom is bonded to the glass substrate, which is cleaned in advance using sonication in ethanol followed by plasma treatment (45~W, 1~min, PDC-001-HP, Harrick Plasma). 

The PDMS microchannel with a height of $h=15\pm1$~\um~consists of two wide parts with transverse dimensions of $L_y=200$~\um~[Fig.~\ref{fig:SI-setup}(b)] and a narrow part with $w=15$~\um~in between. The inlet and outlet holes of the PDMS block are fabricated by a biopsy punch of a diameter of 1.5~mm. The hole fabrication is done by hand. Therefore, although the longitudinal length of the narrow part is accurately fabricated as $L_x=500$~\textmu m [Fig.~\ref{fig:SI-setup}(b)], the longitudinal lengths of the wide parts, $L_{\mathrm{in}}+L_{\mathrm{out}}$, may vary for each device. Approximately, the values of $L_{\mathrm{in}}+L_{\mathrm{out}}$ are estimated to be $\approx 10$~mm using microscope observation. Silicone tubes connect the inlet and outlet holes of the PDMS block to two reservoirs via glass tubes, respectively [Fig.~\ref{fig:SI-setup}(a)]. 

The magnitude of flows in the test section can be controlled by setting the pressure drop $\Delta p$ across the narrow part of the microchannel over the length $L_x$. 
Here, $\Delta p$ is expected to be proportional to $\Delta H$, which can be controlled with the resolution of $1$~\um~by using the $z$ stage [Fig.~\ref{fig:SI-setup}(a)]. 
In the present experiment, residual bubbles in the microchannel must be avoided, otherwise the flow control may fail. For instance, if residual bubbles exist, there remains weak but non-negligible flow even though we set $\Delta H = 0$~\um. To remove the residual bubbles, we use the degassing process of the whole device in the same manner as described in Ref.~\cite{Tsuji2023}.

\subsection{Pressure gradient}\label{sec:SI-pressure-gradient}

The pressure gradient $|\nabla_x p|$ in Eq.~\eqref{eq:flow} is evaluated as follows. 
First, the pressure difference between two reservoirs, $\Delta p_{\all}$, is estimated as  $\Delta p_\all = \rho g \Delta H\approx0.49$~Pa, where $\rho=1\times10^3$ kg/m$^{3}$ is the density of the fluid, $g=9.8$~m/s$^{2}$ is the acceleration of gravity, and $\Delta H=50$~\um~is the water head difference [Fig.~\ref{fig:SI-setup}(a)]. The pressure difference $\Delta p_\all$ acts over the entire flow path, including the narrow and wide parts of the microchannel as well as the silicone tubes. The flow resistance of the tubes can be neglected because of the large diameter of 1~mm. Therefore, the flow resistance of the narrow and wide parts is dominant and is denoted by $R_{\mathrm{narrow}}$ and $R_{\mathrm{wide}}$, respectively. 

The narrow-part flow resistance is estimated as $R_{\mathrm{narrow}}=2.80\times10^{-4}$~Pa$\cdot$s/\textmu m$^{3}$ using the length $L_x=500$~\um, the height $h=15$~\um, and the width $w=15$~\um.
The wide-part flow resistance is estimated as $R_{\mathrm{wide}}=1.87\times10^{-4}$ Pa$\cdot$s/\textmu m$^{3}$ using the length $L_{\mathrm{in}}+L_{\mathrm{out}}\approx 10$~mm, the height $h=15$~\um, and the width $w=200$~\um. 
Thus, the flow resistance at the narrow part accounts for the portion $\alpha=R_{\mathrm{narrow}}/(R_{\mathrm{narrow}}+R_{\mathrm{wide}})$ of the overall flow resistance, and the pressure difference over the narrow part is estimated as $\Delta p =  \alpha \Delta p_\all = 0.294$~Pa. Note that the fabrication (and/or measurement) error of $\delta L=\pm 1$~mm in the length $L_{\mathrm{in}}$ (or $L_{\mathrm{out}}$) results in the uncertainty of $\pm 0.039$~Pa, that is, the pressure difference in the narrow part is considered to be in the range $\Delta p = 0.294\pm0.039$~Pa in our experiments. The error of $\delta L=\pm 1$~mm is reasonable since we fabricate $L_{\mathrm{in}}$ (or $L_{\mathrm{out}}$) by hand. 

\section{Effect of temperature increase in optical trapping}\label{sec:compute-thermoosmosis-thermophoresis-effect}

\begin{figure*}[bt]
    \centering
    \includegraphics[width=1\linewidth]{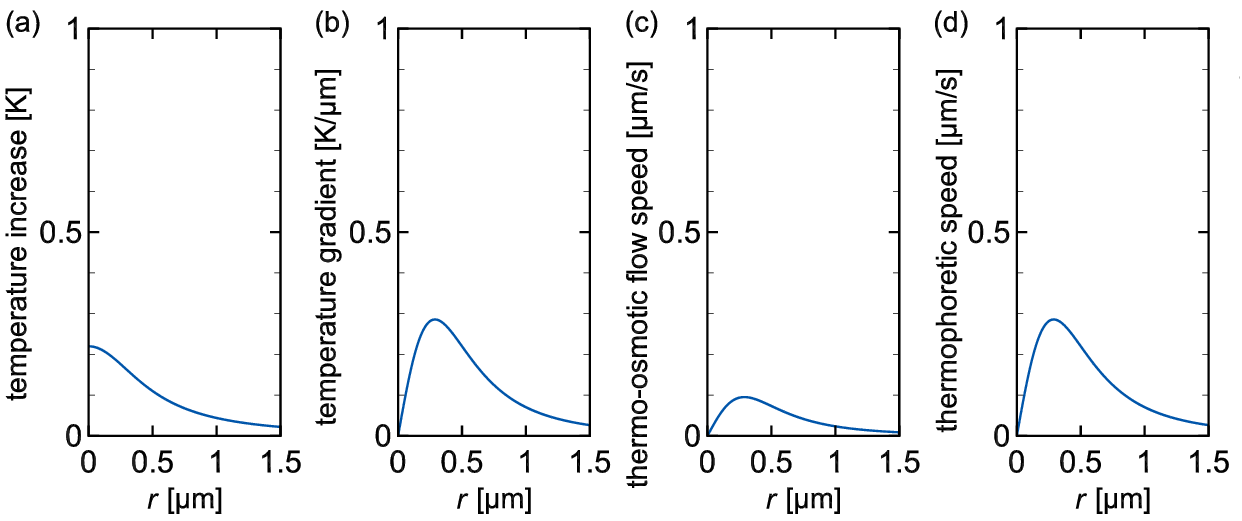}
    \caption{(a) temperature increase $T-T_0$, (b) the magnitude of temperature gradient $|\dd T/\dd r|$, (c) thermo-osmotic flow speed, and (d) thermophoretic speed as a function of the radial distance $r$, for a typical experimental parameter set described in Sec.~\ref{sec:compute-thermoosmosis-thermophoresis-effect}.}
    \label{fig:TO-TP}
\end{figure*}

The temperature increase due to laser heating can be well approximated by a half-Lorentzian curve, as demonstrated in Refs.~\cite{Cordero2009,Tsuji2021,Tsuji2023}, that is, the temperature of fluid $T$ is expressed as a function of the radial direction $r$ as
\begin{align}
T(r)=T_\infty + \frac{\Delta T_{\max}}{1+(r/\sigma)^2}, 
\end{align}
where $r$ is the distance from the laser, $T_{\infty}$ is a room temperature, $\Delta T_{\max}$ is the maximum temperature increase, $\sigma$ is the half width at mid-height of the profile. We try to estimate the effects of temperature increase, thermo-osmosis, and thermophoresis using this model, by estimating $\Delta T_{\max}$ and $\sigma$. 

First, the maximum temperature increase $\Delta T_{\max}$ is estimated as $\Delta T_{\max}=P \times C_{\Delta T/\Delta P} \times C_{\mathrm{duty}}$, where $P\approx24$~mW is the laser power in our experiment, $C_{\Delta T/\Delta P}=22$~K/W is the temperature elevation coefficient for H$_2$O at the wavelength $1064$~nm \cite{Ito2007}, and $C_{\mathrm{duty}}=40~\%$ is the ratio of the trap phase in our experiments. Recall that the laser is not irradiated in the release phase, which is $60~\%$ of the experimental duration. These estimates lead to $\Delta T_{\max} = 0.22$~K at most. Note that heat diffuses in the release phase. High thermal diffusivity of water, $\alpha_{\mathrm{water}}\approx 1\times 10^5$~\textmu m$^2$/s, indicates that the heat diffuses away from the region of interest (say, the size with the order of $1$~\textmu m) within a millisecond. Therefore, $\Delta T_{\max}$ is expected to be much smaller than $0.22$~K in reality. 

Next, we estimate the parameter $\sigma$, which is the effective size of the heated region. The $\sigma$ should be related to the beam waist, and is expected to be larger than the beam waist. Since the magnitude of temperature gradient is $\displaystyle\max_r|\dd T/\dd r|\propto \sigma^{-1}$, smaller $\sigma$ leads to larger $T$. Here, to estimate the upper limit of the temperature gradient, we put $\sigma=w$, where $w\approx 500$~nm is the ideal beam waist near the diffraction limit.

Figure~\ref{fig:TO-TP}(a) and (b) show the temperature increase $T-T_0$ and the magnitude of temperature gradient $|\dd T/\dd r|$, respectively as a function of $r$. The temperature increase is small so that the change of physical parameters, such as viscosity, is negligible. Figure~\ref{fig:TO-TP}(c) and (d) show the magnitude of thermo-osmotic flow speed $v_{\TO}= -\chi\nabla T/T$ \cite{Bregulla2016} and thermophoretic speed $v_{\TP}=-D_T \nabla T$ \cite{Piazza2008}, respectively, where $\chi=1\times10^{-10}$~m$^2$/s \cite{Bregulla2016} (for glass substrate) and $D_T=1$~\textmu m$^2$/s~K for typical colloids are used. The effects of thermo-osmosis and thermophoresis are estimated below $0.5$~\textmu m/s even when we greatly overestimate the temperature increase $\Delta T_{\mathrm{max}}$, and thus they are negligible for our experiment with flow speed $\approx 10$~\textmu m/s.

\section{Evaluation of optical force acting on tracers}\label{sec:optical-force}

In this section, we estimate the optical force acting on tracers for our experimental conditions.
The optical gradient force $\bm{F}=(F_r,\,F_z)$ acting on a spherical tracer particle is given as \cite{Ito2010}
\begin{align}
\bm{F}(r,\,z)=-\nabla U (r,\,z), \quad U(r,\,z) = - U_0 \exp\left(-\frac{r^2}{R_{xy}^2}-\frac{z^2}{R_z^2}\right), \quad U_0 = \frac{n_m \alpha_p}{c} \frac{P}{\pi R_{xy}^2},
\end{align}
where $(r,\,z)$ is the cylinderical coordinate with radial position $r$ and axial position $z$, $U$ is the optical potential, $U_0$ is the potential depth, $R_z(\approx 5R_{xy})$ and $R_{xy}$ are the sizes of beam waist in the $z$ and $r$ directions, respectively, $n_m\approx1.33$ is the refractive index of water, $c\approx3\times10^{8}$~m/s is the speed of light, $\alpha_p$ is the polarizability of the particle, and $P\approx 50$~mW is the laser power. Based on the Rayleigh scattering theory, $U_0$ is estimated as $U_0=4\pi a^3 (n_p^2-n_m^2)/(n_p^2+2n_m^2)$, where $a=d/2$ is the tracer radius and $n_p\approx 1.57$ is the refractive index of the tracer, which is made of polystyrene. Note that the Rayleigh scattering theory is applicable to particle diameters smaller than the wavelength, which is not our case. However, it has been shown that the theory gives a reasonable estimate for the radial component $F_r$, as discussed in \cite{Harada1996}. In the $z$ direction, the theory overestimates the force $F_z$. We use the Rayleigh scattering theory as an approximation for the estimate of $F_r$ below.

\begin{figure*}[bt]
    \centering
    \includegraphics[width=0.6\linewidth]{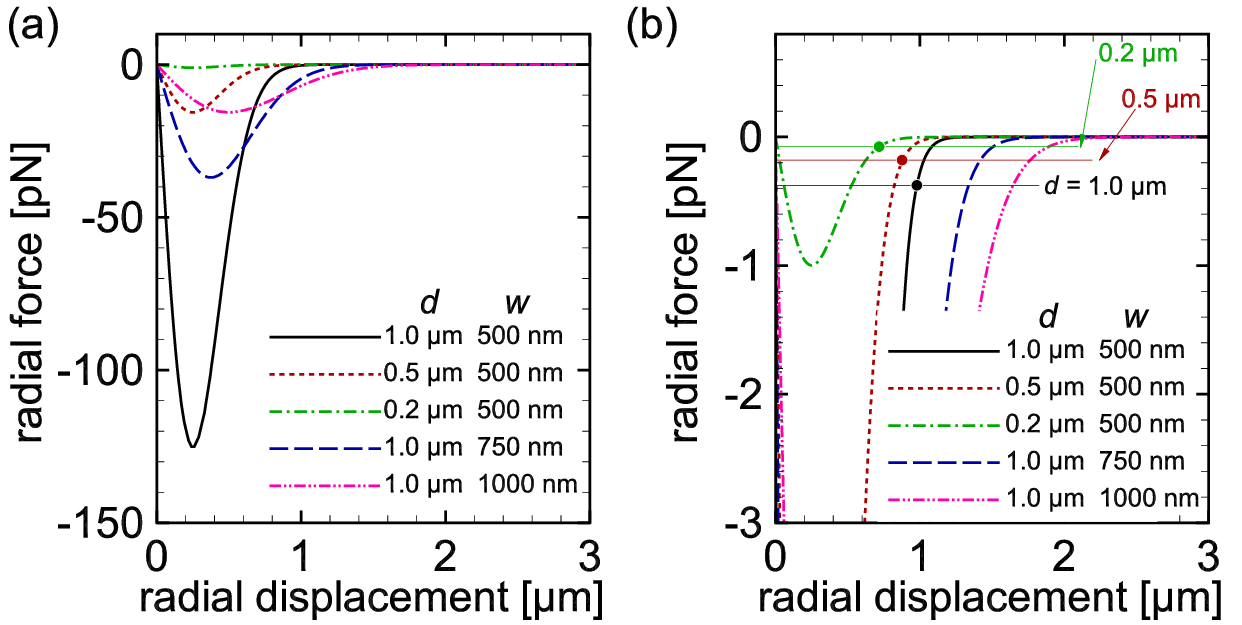}
    \caption{(a) Radial component of the optical force $F_r$ as a function of the displacement $r$ for various values of the tracer diameter $d$ and the beam waist $w$. (b) Magnification near $F_r=0$. Thin-horizontal lines represent the corresponding Stokes drag.}
    \label{fig:optical}
\end{figure*}

Figure~\ref{fig:optical}(a) shows the radial force $F_r$ as the function of the displacement $r$ from the equilibrium position, for the diameter $d=1$~\textmu m, $500$~nm, $200$~nm and the beam waist $w(=R_{xy})=500$~nm, $750$~nm, $1000$~nm. The force decreases rapidly as $d$ becomes small. When $w$ is increased from the near-diffraction-limit condition $w=500$~nm to $1000$~nm, the force magnitude is decreased while the effective range of $r$ is stretched.

To discuss the feasibility of optical trapping under the flow field, we compare $F_r$ with the Stokes drag. Figure~\ref{fig:optical}(b) shows the magnification near $F_r=0$ of panel (a), with the magnitude of the Stokes drag $F_d = -|3\pi d \mu v|$ (minus sign is there to compare with $F_r$) indicated by thin-horizontal lines. Here, $\mu=1\times10^{-3}$~Pa~s is the viscosity and $v$ is the relative speed of the tracer to the fluid. That is, the intersection (indicated by circles) between $F_r$ and the thin-horizontal lines represents the radial range $r_c$ in which we can pull the tracer back to the initial position under the flow field. Here, we set $v=40$~\textmu m/s because the tracer speed can reach this value due to the Brownian motion (see Fig.~\ref{fig:histgram-selected}). The case of $d=1$~\textmu m/s leads to $r_c\approx 1$~\textmu m, and, as $d$ becomes smaller, $r_c$ becomes smaller. When $w$ becomes larger, $r_c$ becomes larger, so one may think a larger $w$ would be better. However, for larger $w$ (i.e., loose trap), the force balance in the $z$ direction becomes worse, since the optical scattering force in the $z$ direction can overtake the gradient force, resulting in pushing away the tracer in the beam propagation direction. Therefore, a tightly-focused beam (smaller $w$) with larger $d$ is better for optical trapping. 

The typical tracer displacement $r$ can be estimated as $V \Trelease\approx 0.6$~\textmu m, where $V=10$~\textmu m/s is taken from the maximum flow speed in the present paper (see Fig.~\ref{fig:profile}). This is well comparable to $r_c$ discussed above, and thus validates our estimate of optical trapping feasibility. 

\section{Movies}\label{sec:movie}
Movie file ``Fig2b.avi" corresponds to the snapshots of Fig.~\ref{fig:protocol}(b).

Movie files ``Fig8a\_z2um.gif" and ``Fig8b\_z4um.gif" correspond to Figs.~\ref{fig:application-results}(a) and \ref{fig:application-results}(b), respectively. 
The large particle at the center is a light-absorbing particle with $d=3.9$~\textmu m. The heating laser is irradiated at the center.

\section{The uncertainty of the initial position}\label{sec:initial}

\begin{figure*}[bt]
    \centering
    \includegraphics[width=0.35\linewidth]{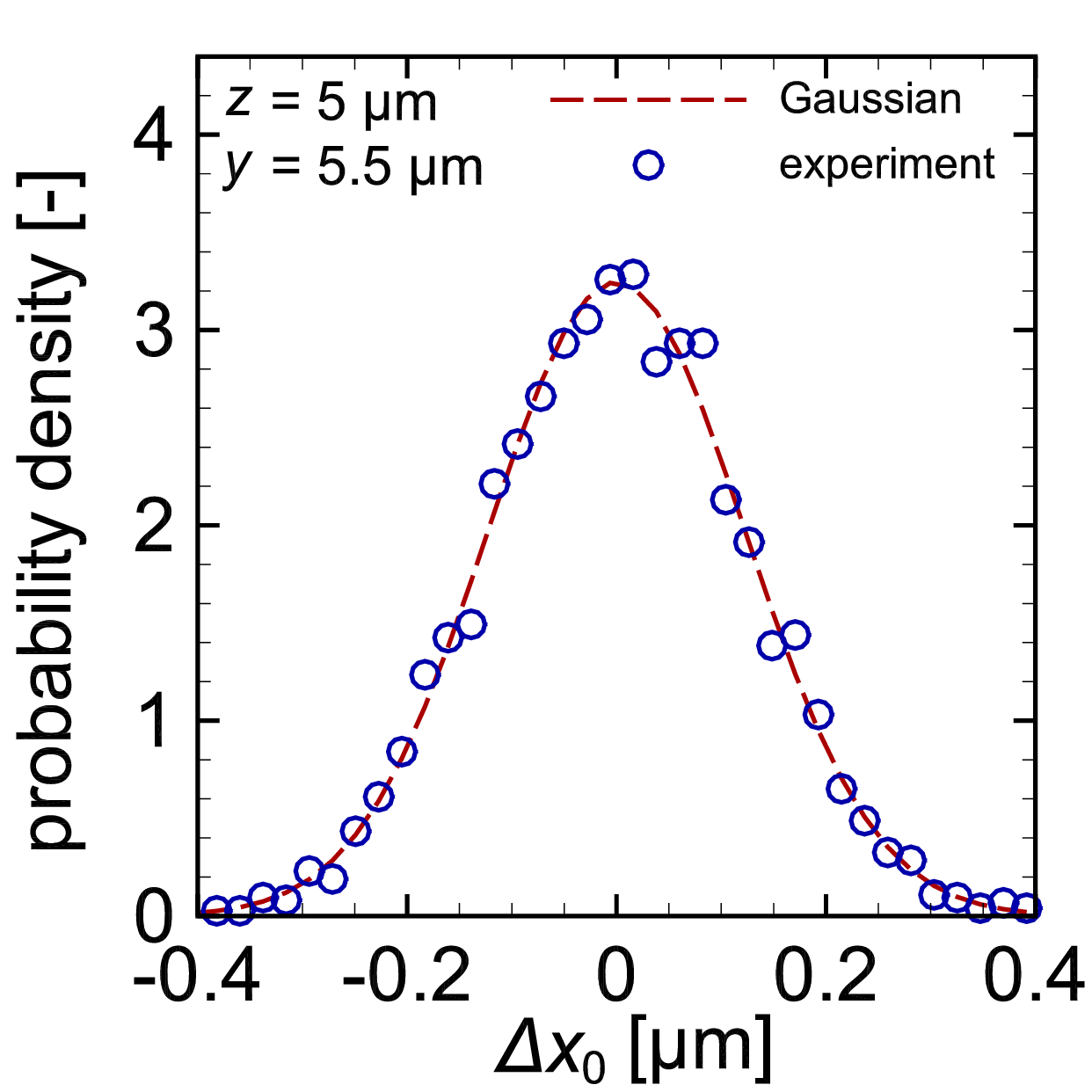}
    \caption{Probability density of $\Delta x_0$ for the data $z=5$~\textmu m and $y=5.5$~\textmu m. Symbols and a dashed curve represent experimental results and their Gaussian fitting, respectively.}
    \label{fig:initial}
\end{figure*}

Figure~\ref{fig:initial} shows the probability distribution of $\Delta x_0$, which indicates the displacement of the tracer's initial position from their ensemble average, where the ``initial position" in this context means the position in the first frame of a release phase. The distribution is well fitted by the Gaussian distribution, as expected, and the standard deviation of the Gaussian distribution is obtained as $123$~nm. This is a quantitative measure of the uncertainty of the initial position, as discussed in the main text.

\section{Probability density of tracer velocity for all measurement positions}\label{sec:SI-probability-density}
\begin{figure*}[bt]
    \centering
    \includegraphics[width=1\linewidth]{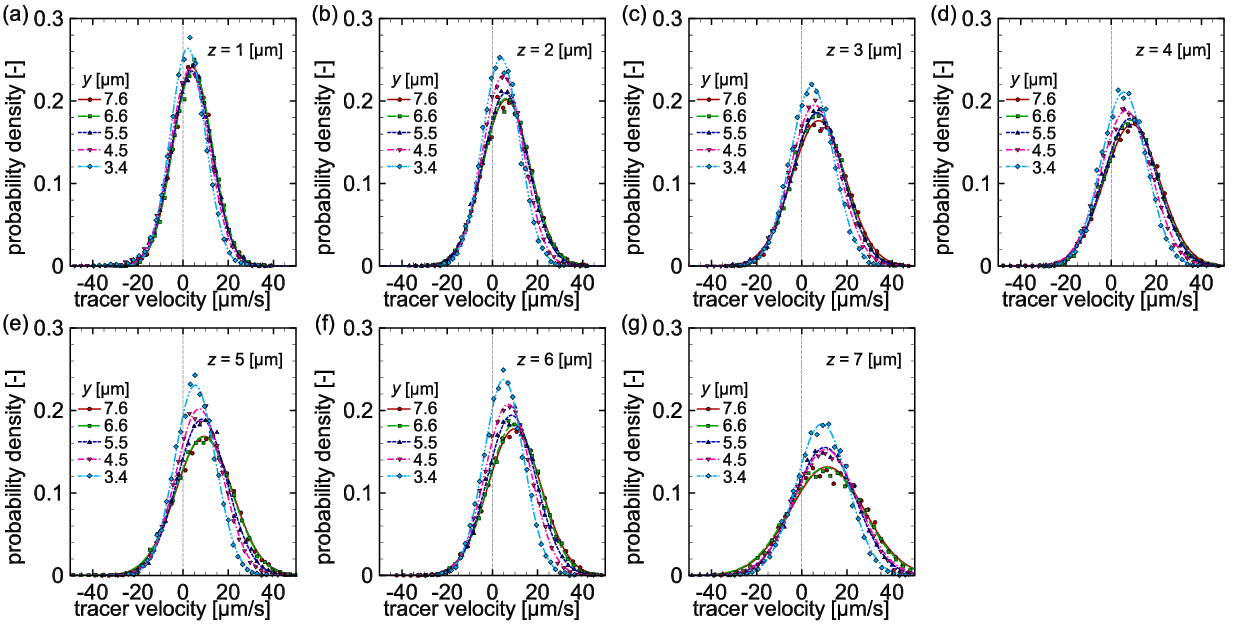}
    \caption{Probability density of tracer velocity $v_x$ at 
    (a) $z=1$~\textmu m, (b) $2$~\textmu m, (c) $3$~\textmu m, (d) $4$~\textmu m, (e) $5$~\textmu m, (f) $6$~\textmu m, and (g) $7$~\textmu m for $y=7.6$~\textmu m (red-circle), $6.6$~\textmu m (green-square), $5.5$~\textmu m (blue-triangle), $4.5$~\textmu m (magenta-lower-triangle), and $3.4$~\textmu m (cyan-diamond). Curves show the corresponding Gaussian fits. The sample size is $N>6600$ for all cases. The results for $z=2$ and $5$~\textmu m are the same as those presented in Fig.~\ref{fig:histgram-selected} of the main text.}
    \label{fig:histgram-all}
\end{figure*}

Figure~\ref{fig:histgram-all} shows the corresponding data of Fig.~\ref{fig:histgram-selected} for all $z$ positions that are measured. The data at $z=4$ [panel (d)] and $7$~\textmu m [panel (g)] are not used in the analysis of Fig.~\ref{fig:profile} because of the extraordinary variance, i.e., the width of the profiles. More specifically, the variances observed in panels (d) and (g) are systematically larger than those in the other panels [i.e., (a), (b), (c), (e), and (f)], and deviate from the trend observed at the other measurement positions. These inconsistencies possibly indicate some failure of the control parameters (e.g., flow speed, measurement position, etc.) and/or errors in tracking analysis. 

\section{Result of PIV analysis}\label{sec:piv}

\begin{figure*}[bt]
    \centering
    \includegraphics[width=0.35\linewidth]{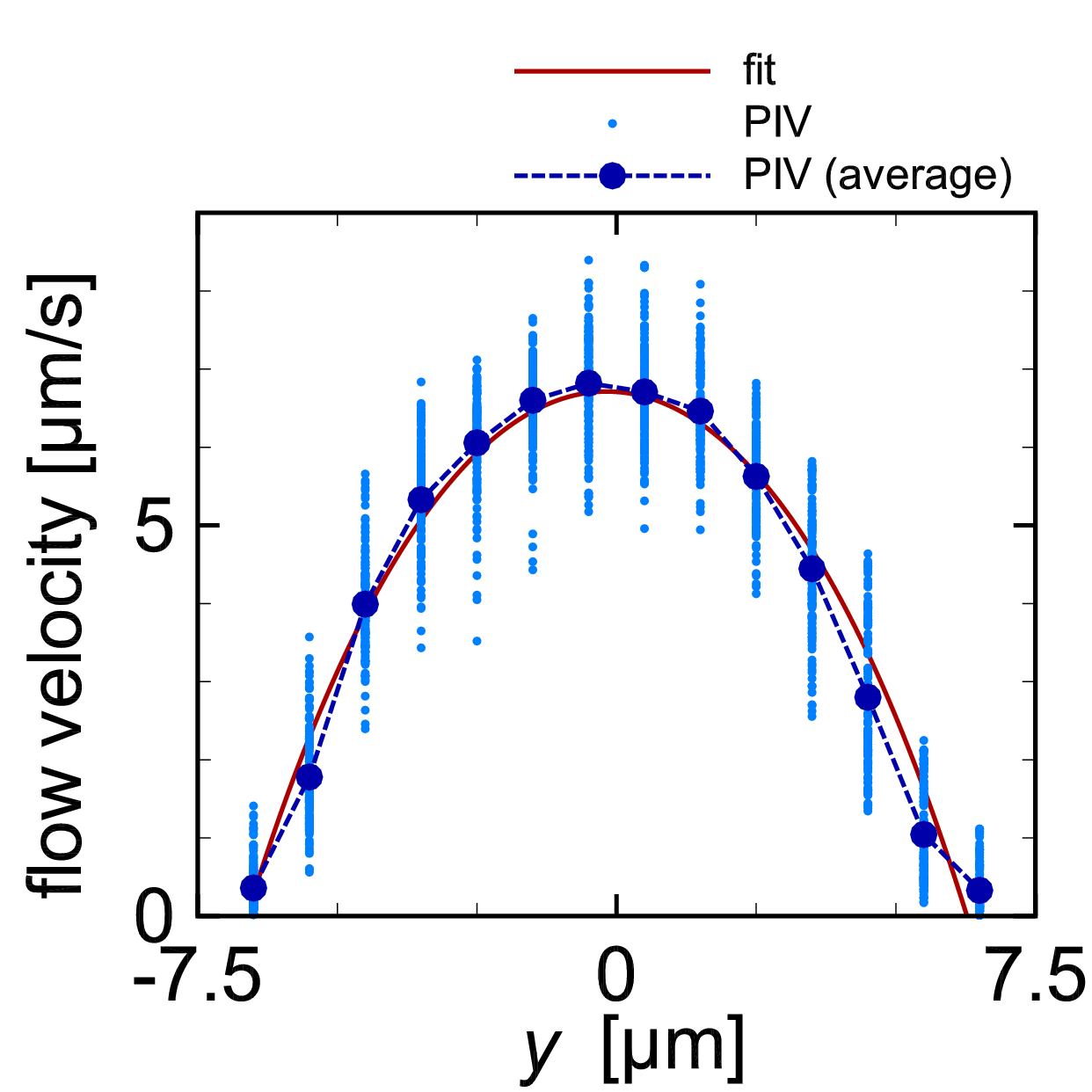}
    \caption{Result of PIV for the same test case of pressure-driven flows in a straight microchannel with the square cross-section. Small and large symbols indicate the result at each $x$ and the average over the whole $x$, respectively.}
    \label{fig:piv}
\end{figure*}

In this section, we show the result of PIV for the same test case of pressure-driven flows in a straight microchannel with the square cross-section. The same pressure difference as the ot-PTV experiment, $\Delta p_\all \approx0.49$~Pa, is applied (see Sec.~\ref{sec:SI-pressure-gradient}). The $z$ position of the measurement is set to $7.5$~\textmu m. The image acquisition area has the lengths $140$~\textmu m ($2160$ pixels) and $18$~\textmu m ($275$ pixels) in the $x$ and $y$ direction, respectively. 

Because of the slow speed of the flow, the frame rate is 10 frames per second and the total acquisition time is $10$~s (i.e., $100$ frames). The number density of tracers is set to $0.02$~wt\%. This density is too low to carry out PIV. However, the sample solution with a higher tracer density is expensive, and thus is avoided. To compensate for the low tracer density, we carried out the same experiments $30$ times and took the average over the all runs. In addition, we carry out the time average over the all frames.

Since the present flow is uniform in the $x$ direction, we can further take the spatial average over the whole $x$ direction. Using the data for different $x$ positions, we can obtain $N_{x\text{-ave}}(>130)$ profiles of the flow velocity $v_x$ as a function of $y$. Figure~\ref{fig:piv} shows the result of PIV. The data at each $x$ are plotted in Fig~\ref{fig:piv} using small symbols, and the spatial average over the whole $x$ is shown by large symbols. The data at each $x$ are rather scattered, possibly due to the diffusion of the tracers. In fact, the frame interval of $\Delta t = 0.1$~s for the present PIV and the diffusion coefficient $D=0.63$~\textmu m$^2$/s (see Sec.~\ref{sec:prob}) lead to the standard deviation $\sqrt{2 D \Delta t}/\Delta t = 3.5$~\textmu m/s. The red solid curve is the curve fit using Eq.~\eqref{eq:flow}, with $\Delta y = -0.9$~\textmu m, $\Delta p=0.24$~Pa, where the channel width $w=12.9$~\textmu m is also used as the fitting parameter in PIV analysis. The fitted parameter $w\approx 13$~\textmu m is smaller than the specification ($w=15$~\textmu m) due to the fabrication uncertainty.

The fitting seems to work satisfactorily in Fig.~\ref{fig:piv} even though the acquisition time is only $10$~s. However, this success is basically due to the average over $x$, which is allowed by the $x$-uniformity of the present test case. In general, when the spatial average is not available, a longer time average will be required [say, $N_{x\text{-ave}}(>130)$ times longer], leading to the experimental time of several minutes or more.


\end{document}